\documentstyle[11pt,epsfig]{article}
\begin{document}
\title{\bf{Distribution of consecutive waves in the sandpile model on
the Sierpinski gasket}}
\author{F. Daerden$^{1}$, V. B. Priezzhev$^{2}$, C. Vanderzande$^{1}$ \\
$^{1}$ {\it Departement WNI, Limburgs Universitair Centrum, 3590
Diepenbeek, Belgium} \\
$^{2}$ {\it Bogolubov Laboratory of Theoretical Physics, Joint Institute for
Nuclear Research,}\\ {\it Dubna, 141980 Russia}
}
\maketitle

\ \\
\ \\
\ \\
\begin{abstract}
The scaling properties of waves of topplings of the sandpile model
on a Sierpinski gasket are investigated. The exponent describing the
asymptotics of the distribution of last waves in an avalanche is
found. Predictions for scaling exponents in the forward and backward
conditional probabilities for two consecutive waves are given. All
predictions were examined by numerical simulations and were found
to be in a reasonable agreement with obtained data.
\end{abstract}
\newpage
\section{Introduction}
Sandpile models form the paradigmatic examples of the concept of
self organised criticality (SOC) \cite{Bak1,Bak2}. This is the
phenomenon in which a slowly driven systems with many degrees of
freedom evolves spontaneously into a critical state,
characterised by
long range correlations in space and time.

In the past decade much progress has been made in the
theoretical understanding of sandpile
models. This is especially true for the
Bak-Tang-Wiesenfeld (BTW) model \cite{Bak1,Bak2},
where, following the original work of Dhar \cite{Dhar},
a mathematical formalism was developped [4--8]
that allows an exact calculation of several properties
of the model such as height probabilities \cite{DharM, Prh},
the upper critical dimension \cite{DcUp} and so on.

Despite all this work, it has however not been possible
yet to give a full and exact characterisation of the
scaling properties of the avalanches in the BTW-model.
In recent years it has become increasingly clear that, especially
in two dimensions, avalanches are to be described
by a full multifractal set of scaling exponents \cite{Mario1,Mario2}.
This spectrum of exponents has been calculated with
high numerical precision, but at this moment there is
no clue how it can be determined by an analytical approach.

Avalanches can be decomposed into simpler objects called
waves \cite{Waves}. The probability distribution of
waves seems to obey simple scaling \cite{KLGB} and
the exponent describing that scaling is known exactly,
both for the general wave \cite{Waves,KLGB} and for
the last wave of each avalanche\cite{LastW}. Since in dimensions $d \geq d_{c}=4$,
multiple topplings are extremely rare, it is
to be expected that in these situations, wave and avalanche
statistics obey the same scaling properties.

More recently, the distribution of two consecutive waves has
received considerable attention \cite{PB,KP}.
The conditional probability that the $k+1$-th wave has size
$s_{k+1}$ given that the previous wave had size $s_{k}$,
$P(s_{k+1}|s_{k})$, is the first quantity to study when
one is interested in correlation effects in waves.
It are these correlations that make waves
and avalanches different. Paczuski and Boettcher \cite{PB}
proposed, on the basis of extensive simulations, that
$P(s_{k+1}|s_{k})$ has a scaling form
\begin{eqnarray}
P(s_{k+1}|s_{k}) \sim s_{k+1}^{-\beta}F(\frac{s_{k+1}}{s_{k}})
\label{1}
\end{eqnarray}
where for large $x$, $F(x) \sim x^{-r}$, while $F(x) \to
\mbox{constant}$ for $x \to 0$. Numerical
estimates for these exponents in $d=2$ are $\beta \approx 3/4$,
$r\approx 1/2$. At this moment, no exact values for these exponents
are known.
In a very recent work, Hu {\it et al.} \cite{Hu} study the
`backward' conditional probability $P(s_{k}|s_{k+1})$ and
find that it obeys a similar scaling law
\begin{eqnarray}
P(s_{k}|s_{k+1}) \sim s_{k}^{-\bar{\beta}}\bar{F}(\frac{s_{k}}{s_{k+1}})
\label{2}
\end{eqnarray}
where for large $x$, $\bar{F}(x) \sim x^{-\bar{r}}$, while $\bar{F}(x) \to
\mbox{constant}$ for $x \to 0$. These
authors give arguments that show that
\begin{eqnarray}
\bar{\beta}+\bar{r}=\tau_{lw}
\label{3}
\end{eqnarray}
where $\tau_{lw}$ is the scaling exponent describing
the size distribution of the last wave. The relation (\ref{3})
is consistent with the numerical data for the square lattice
where both Euclidean and fractal dimensions of waves are $2$.
At the same time, it is desirable to get an independent
verification of (\ref{3}) using lattices of different dimensions.

In the present paper we study the properties of waves on the
Sierpinski gasket, continuing previous work \cite{FD2}.
Using the methods of analysis introduced in \cite{Mario1},
we obtain precise numerical estimates for the exponents
$\tau_{lw},\beta,r,\bar{\beta}$ and $\bar{r}$. Also in this
case equation (\ref{3}) seems to be well satisfied
which indicates that short time correlations
in waves admit an analytical treatment.
\section{The sandpile model on a Sierpinski gasket}
The BTW sandpile model can be defined on any graph, but for
definiteness
we will introduce it in the context of the Sierpinski gasket (see figure 1).
Each vertex (apart from the three boundary sites)
of this graph has four nearest
neighbours. To each such vertex $i$ we associate a height variable
$z_{i}$ which can take on any positive integer number. We also
introduce a critical height $z_{c}$, which we will take equal to four
for all vertices. The number of sites in the lattice, $N$, is
trivially related to the number of iterations $n$ used in
constructing the fractal. The dynamics is defined as follows.
On a very slow time scale we drop sand at a randomly selected
site and thereby increase the height variable by one: $z_{i}
\to z_{i}+1$. When at a given site, $z_{i} > z_{c}$, that site
becomes {\it unstable} and {\it topples}
\begin{eqnarray*}
z_{j} \to z_{j} - \Delta_{i,j}
\end{eqnarray*}
where
\begin{eqnarray}
\Delta_{i,j}= \begin{array} {rl}
 4 & i=j\\
-1 & \mbox{if $i$ and $j$ are nearest neighbours}\\
0 & \mbox{otherwise} \end{array}
\label{4}
\end{eqnarray}
Through toppling, neighbouring sites can become unstable, topple
themselves, create new unstable sites, and so on. This {\it avalanche}
of topplings proceeds on a very fast time scale and no new grains of
sand are added before the avalanche is over. Sand can leave the system
when a boundary site topples. An avalanche is over when all
sites are stable again.

\begin{figure}
  \centerline{ \epsfig{figure=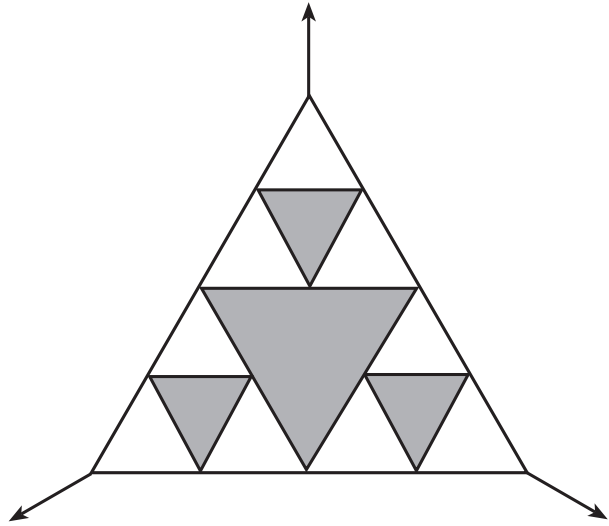,width=8cm}}
\caption{Sierpinski gasket with $n=2$. The arrows indicate the
three sites where sand can leave the system.}
\label{figure 1}
\end{figure}

For further reference it is also necessary to introduce the
matrix $G$, called the lattice Green function, which is the inverse of $\Delta$.

It is not difficult to see that the order in which unstable
sites are toppled does not influence the stable configuration
which is obtained when the avalanche is over. This Abelian
nature of the sandpile model allows the introduction of
the concept of {\it waves} \cite{Waves} which are introduced in
the following way. Suppose that an avalanche starts at a site
$i_{0}$ and that after a few topplings $i_{0}$ becomes unstable again.
One can then forbid $i_{0}$ to topple again and continue with
the toppling of other unstable sites untill all of them are
stable. It is easy to show that in such a sequence all sites topple at
most once. This set of topplings is called {\it the first wave}.
Next, we topple the site $i_{0}$ for the second time. If after some
topplings it becomes unstable again, we keep it fixed, and topple all
the other unstable sites. This set of topplings constitutes the
second wave. We continue in this way untill we finally reach
a stable configuration. We can in this way decompose any avalanche
in a set of waves. The probability distribution $P_{w}(s)$ that
an arbitrary wave involves $s$ topplings obeys simple scaling
\cite{KLGB}
\begin{eqnarray}
P_{w}(s) \approx s^{-\tau_{w}}
\label{5}
\end{eqnarray}
(For the moment we neglect finite size effects which will be taken
into account in section 4.)

Because in a wave, sites topple at most once, waves are simpler objects to
analyse than avalanches.
Without going into details we summarize the following important
properties of waves \cite{Waves}
\begin{itemize}
\item There is a one-to-one correspondance between waves and the
two-rooted spanning trees on a graph which consists of the
Sierpinski gasket and one extra site called the sink. The sink
is connected with two edges to each of the three boundary sites
of the Sierpinski gasket.
\item The element $G_{ij}$ of the Green function is given by the ratio
of the number of two-rooted spanning trees (in which $i$ and $j$ are
in the same subtree) to the number of one-rooted spanning trees.
Moreover $G_{ij}$ is also equal to the expected number of topplings
at site $j$ when a grain of sand was dropped at site $i$, which
is proportional to the probability that a wave started at $i$ reaches
$j$.
\end{itemize}
From these results, it follows that $P_{w}(R)=dG(R)/dR$ where $R$
is the linear size of the wave. The asymptotic behaviour of the Green function
on an arbitrary lattice is
\begin{eqnarray*}
G(R) \sim R^{d_{w}-d_{f}}
\end{eqnarray*}
where $d_{f}$ is the fractal dimension of the lattice, and $d_{w}$
the dimension of a random walk on the lattice. Therefore,
$P_{w}(R) \sim R^{d_{w}-d_{f}-1}$. From the definition
of fractal dimension, $s \sim R^{d_{f}}$ we finally obtain
\begin{eqnarray}
P_{w}(s) \sim s^{d_{w}/d_{f}-2}
\label{6}
\end{eqnarray}
so that
\begin{eqnarray}
\tau_{w}=2 - \frac{d_{w}}{d_{f}}
\label{7}
\end{eqnarray}
a result first derived in \cite{FD2}. For the particular case of
the Sierpinski gasket, $d_{f}=\log{3}/\log{2}, d_{w}=\log{5}/\log{2}$,
so that $\tau_{w}=\log{(9/5)}/\log{3} \approx 0.535$, a result which
is nicely consisted with the available numerical data.

The properties of the last wave in a given avalanche are of special
interest for us (see section 3). The probability distribution
$P_{lw}(s_{lw})$ that
the last wave has $s_{lw}$ topplings obeys the scaling law
\cite{KLGB}
\begin{eqnarray}
P_{lw}(s_{lw}) \sim s_{lw}^{-\tau_{lw}}
\label{8}
\end{eqnarray}
From the definition of waves it follows that the last wave
has the property that the site $i_{0}$ is on the boundary
of the wave. Let us denote by $d_{B}$ the fractal dimension
of the boundary of an arbitrary wave.
The number of points on the boundary of a wave of size $s$ is then of
order $s^{d_{B}/d_{f}}$. Hence,
the probability that
a given site is on the boundary of a wave of size $s$ is
proportional to $s^{d_{B}/d_{f}-1}$, which should
also be proportional to $P_{lw}(s_{lw})/P_{w}(s)$. Using (6)
we immediately obtain
\begin{eqnarray}
\tau_{lw}=3 - \frac{d_{w}+d_{B}}{d_{f}}
\label{9}
\end{eqnarray}
In two dimensions, $d_{w}=2$ and $d_{B}=z=5/4$ where $z$ is
the fractal dimension of the chemical path on a spanning tree.
One thus finds $\tau_{lw}=11/8$ \cite{LastW}.
On a fractal the relation between $d_{B}$ and $z$ may be more
complicated. In fact, it was shown in \cite{DharDhar} that for
a deterministic fractal
\begin{eqnarray}
d_{B}=z-d_{w}+d_{f}
\label{10}
\end{eqnarray}
so that, from (\ref{9}) we obtain
\begin{eqnarray}
\tau_{lw}= 2 - \frac{z}{d_{f}}
\end{eqnarray}
The exponent $z$ on the Sierpinski gasket was also calculated in
\cite{DharDhar} with the result \newline $z=\log{[(20+\sqrt{205})/15]}/\log{2}$.
Thus one finally obtains for the case of the Sierpinski gasket
\begin{eqnarray}
\tau_{lw}=-\frac{\log{[(20+\sqrt{205})/135]}}{\log{3}} \approx 1.247
\label{11}
\end{eqnarray}
In section 4, we will present numerical estimates for $\tau_{lw}$ that are fully
consistent with this prediction \footnote{The derivation of
$\tau_{lw}$ in \cite{FD2} used implicitly that the graph is
selfdual, which is correct for the square lattice but not for
the Sierpinski gasket. This error was pointed out by one
of us (VBP). The correct result is that given in (\ref{11}).}.
\section{Distribution of consecutive waves}
In order to characterise the statistical properties of waves more
completely, it is necessary to go beyond the description on the
basis of the distribution $P_{w}(s)$ only. Following the work of
Paczuski and Boettcher \cite{PB} we now turn to a study of
the conditional probability $P(s_{k+1}|s_{k})$ where
$s_{k}$ is the size of the $k$-th wave. If the size of
consecutive waves is a Markov process, this conditional probability
is sufficient to describe the evolution of wave sizes.
In figure 2.a, we show our data for this conditional probability
for Sierpinski gaskets with $n=9$ ($N=29526$). All our data were obtained by studying
at least $1000 \times N$ avalanches.
The figure shows the best fit of our data to the scaling form
(\ref{1}) proposed by Paczuski and Boettcher. Unfortunately, it
is not possible to obtain very accurate estimates for the exponents
$\beta$ and $r$ in this way. We will come back to this issue in
the next section.

\begin{figure}
  \centerline{ \epsfig{figure=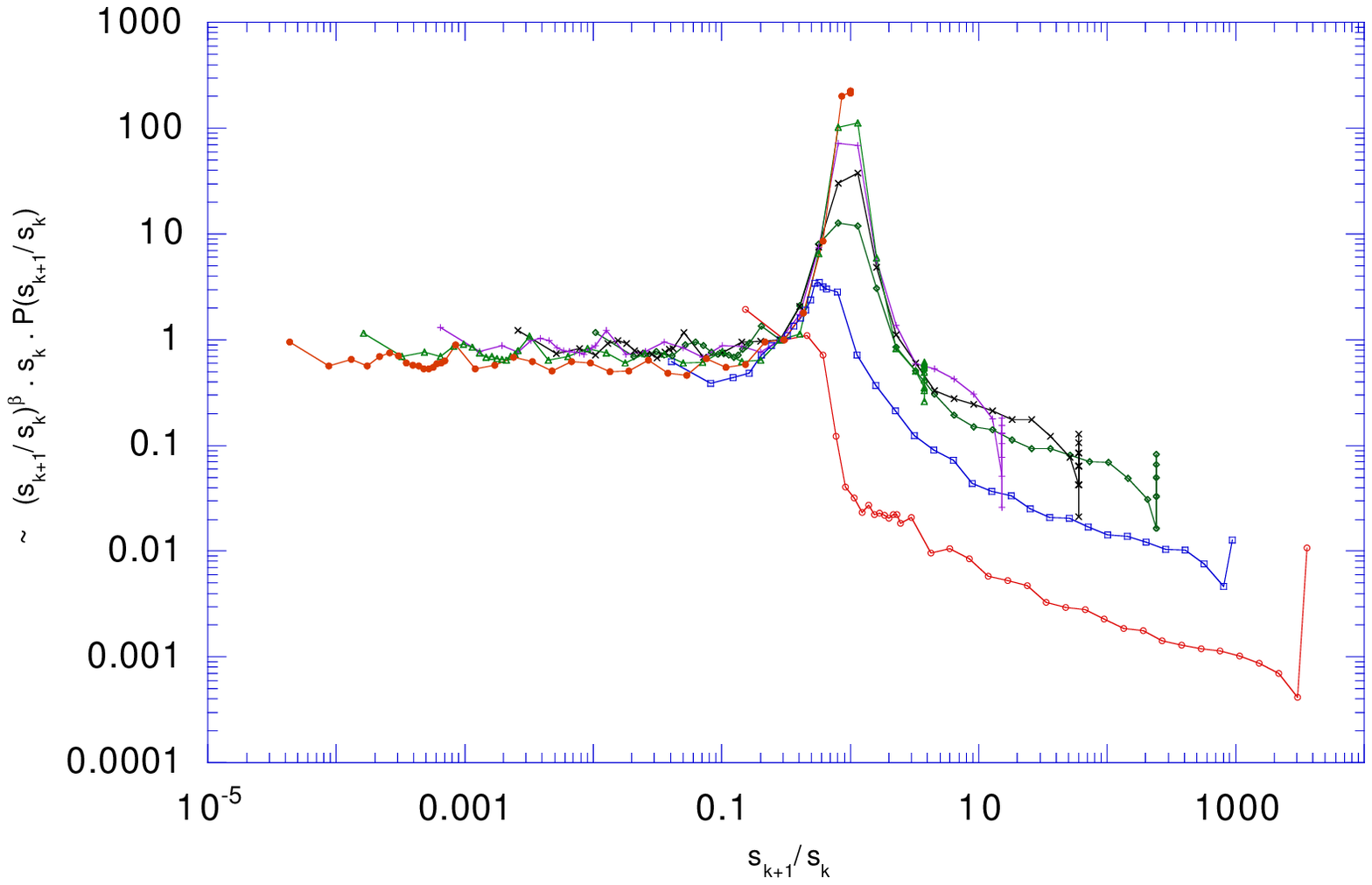,width=14cm}}
  \centerline{ \epsfig{figure=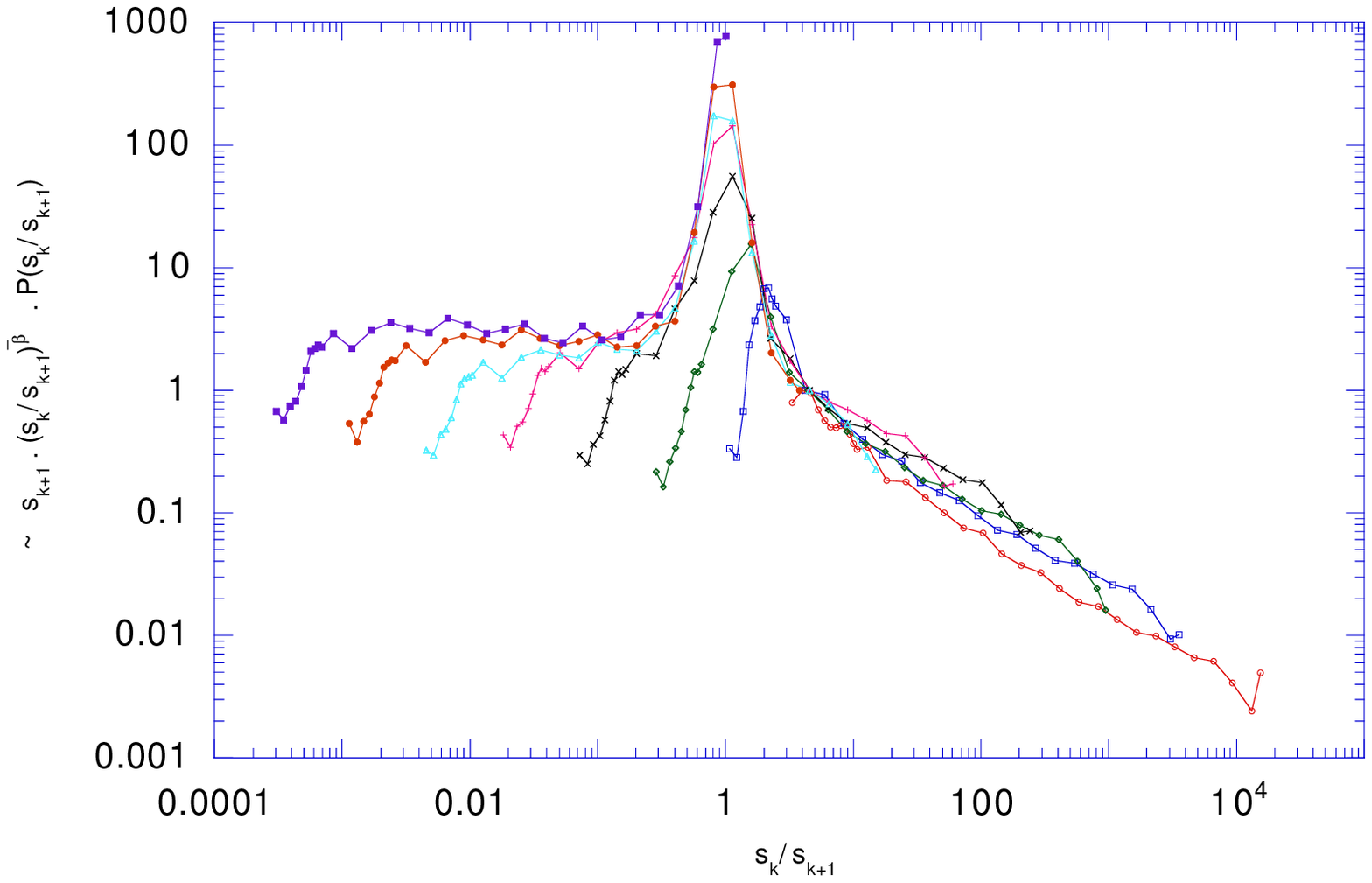,width=14cm}}
\caption{Scaled conditional probabilities $P(s_{k+1}|s_{k})$ (a)
and $P(s_{k}|s_{k+1})$ (b). The results were obtained for a Sierpinski
gasket with $n=9$.}
\label{figure 2}
\end{figure}

Recently, it was pointed out that the `backward' conditional
probability
$P(s_{k}|s_{k+1})$ is also of interest because it is possible
to relate the exponents $\bar{\beta}$ and $\bar{r}$ (see (\ref{2}))
to $\tau_{lw}$.
In figure 2.b we present our data for this quantity, again for the case $n=9$.

We now repeat briefly the argument given in
\cite{Hu}.
We begin by rewriting (\ref{2}) in a normalised form
\begin{eqnarray}
P(s_{k}|s_{k+1}) \sim \left(\frac{s_{k}}{s_{k+1}}\right)^{-\bar{\beta}}
\bar{F}\left(\frac{s_{k}}{s_{k+1}}\right) \ s_{k+1}^{-1}
\label{12}
\end{eqnarray}

Let's next consider the situation in which $s_{k} \gg s_{k+1}$ so that
the argument of $\bar{F}$ in (\ref{12}) is large.
In that case it must be so that the $(k+1)$-th wave has
a non-empty intersection with the boundary of the previous
wave (see figure 3.a). Indeed, assume the opposite so that the $(k+1)$-th
wave covers a small region inside the much bigger $k$-th wave (figure
3.b).
But such a situation is forbidden, since all sites inside
the $k$-th wave return to their original height after the
wave has passed. The $(k+1)$-th wave must therefore follow
the motion of the previous wave untill it hits the boundary
of the $k$-th wave from where it can follow a different evolution.
Therefore, the situation of figure 3.b cannot occur.

\begin{figure}
  \centerline{ \epsfig{figure=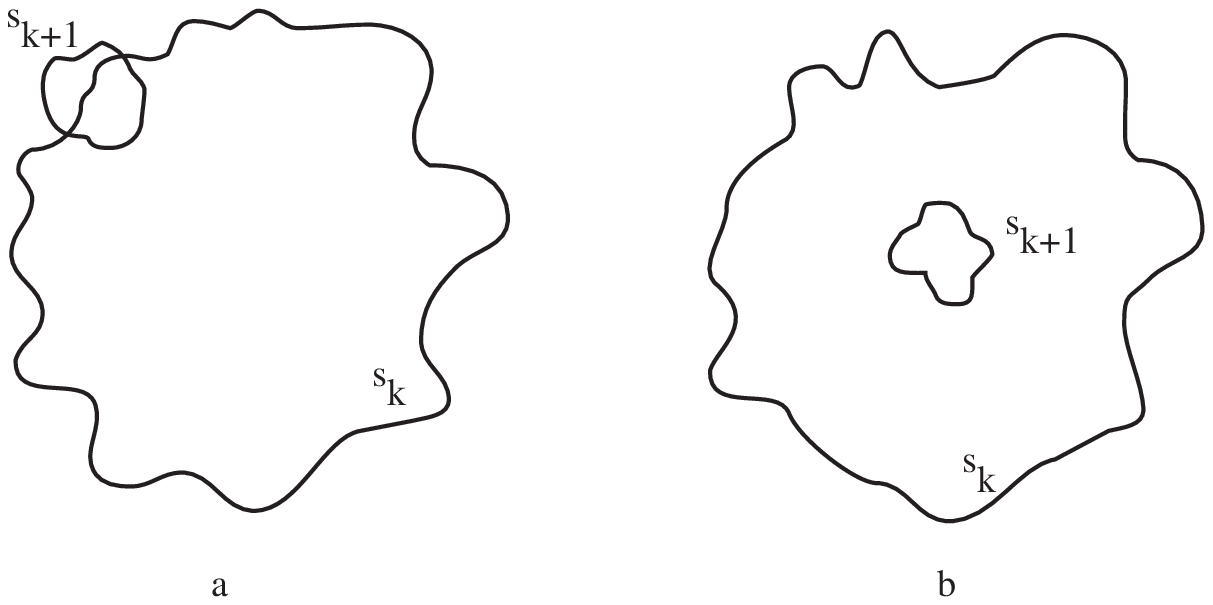,width=14cm}}
\caption{When $s_{k+1} \ll s_{k}$ the $k+1$ the wave must have
an intersection with the boundary of the $k$-th wave as shown
in a. The situation in b cannot occur.}

\label{figure 3}
\end{figure}

Then, consider figure 3.a on a coarse grained scale by performing
a rescaling of the order of $R_{k+1}$, the linear size of the
$k+1$-th wave. On that scale, the geometry of figure 3.a resembles
that of last waves with $s_{k+1}$ playing the role of the origin
of the avalanche, and $s_{k}$ that of the last wave.
Hence we arrive at the conclusion that in the limit $s_{k} \gg
s_{k+1}$, the distribution of $s_{k}$ coincides with that of
the last wave. From (\ref{12}), (\ref{8}) and the asymptotic behaviour of
$\bar{F}$, the equality (\ref{3}) then follows.

Once this result has been obtained it is possible to obtain
also a relation for the exponents $\beta$ and $r$ that appear in the scaling
form (\ref{1}). The joint distribution $P(s_{k},s_{k+1})$ can
be written in two ways using either the forward or backward
conditional probability
\begin{eqnarray*}
P(s_{k},s_{k+1})&=&P(s_{k}|s_{k+1}) P_{w}(s_{k+1}) \\
&=&P(s_{k+1}|s_{k}) P_{w}(s_{k})
\end{eqnarray*}
We then insert (\ref{12}), (\ref{5}) and a properly normalised
version of (\ref{1}) and get
\begin{eqnarray}
\left(\frac{s_{k}}{s_{k+1}}\right)^{-\bar{\beta}}
\bar{F}\left(\frac{s_{k}}{s_{k+1}}\right) \ s_{k+1}^{-1-\tau_{w}}
 \sim
\left(\frac{s_{k+1}}{s_{k}}\right)^{-\beta}
F\left(\frac{s_{k+1}}{s_{k}}\right) \ s_{k}^{-1-\tau_{w}}
\label{extra}
\end{eqnarray}
In the case $s_{k} \gg s_{k+1}$ we insert the proper limiting
behaviours of the functions $F$ and $\bar{F}$, and immediately
obtain, using (\ref{3})
\begin{eqnarray}
\beta=1+ \tau_{w}-\tau_{lw}
\label{14}
\end{eqnarray}

A final equality between exponents can be obtained by
investigating the limit $s_{k} \ll s_{k+1}$ in (\ref{extra}).
Inserting the appropriate scaling behaviours one obtains
\begin{eqnarray}
\beta+r=1+\tau_{w}-\bar{\beta}
\label{extra.1}
\end{eqnarray}

In $d=2$, (\ref{14}) leads to the predictions $\beta=5/8$ and
$\bar{\beta}+\bar{r}=11/8$. The value of $\beta$ is not too far
from the numerical estimate  $\beta \approx 3/4$ reported
by \cite{PB}, while in \cite{Hu} numerical evidence is
presented that is in agreement with the prediction (\ref{3}).
In the following section, we investigate the situation
on the Sierpinski gasket.

\section{Numerical results}
In order to analyse our data we have used the method introduced
in \cite{Mario1}, in which one investigates the moments $\langle
s^{q}\rangle$ of the distribution $P(s,L)$,
where we now explicitly take into
account the size $L$ of the system.
In our case, $L=2^{n}$. If $P(s,L)$ has a simple
scaling form
\begin{eqnarray}
P(s,L) \sim s^{-\tau} H(s/L^{d_{f}})
\label{15}
\end{eqnarray}
these moments should be proportional to simple powers of
$L$, $\langle s^{q}\rangle \sim L^{\sigma(q)}$ where
$\sigma(q)=d_{f}(1-\tau+q)$ for $q > \tau-1$, and
$\sigma(q)=0$ for $q \leq \tau-1$.

In principle, an analysis of the moments is most instructive
when one is interested in the presence of possible multifractal
scaling (instead of simple scaling). In that case, the function
$\sigma(q)$ becomes nonlinear.
However, even in the absence of multifractality, this kind
of analysis has many advantages.
In the case of Sierpinski gasket, the probability distributions for
the size of avalanches, waves and last waves show strong oscillations
superposed on the pure power laws
(see the
figures 2,4 and 5 in \cite{FD2})
. This is a consequence of the
discrete scale invariance of the system. These oscillations
make the determination of the scaling exponents a hard task.
The moments $\langle s^{q} \rangle$ have the big advantage
that they are averages over the distribution and hence
the effects of the oscillations almost completely disappear.
If one then assumes simple scaling, as we know is correct
for waves \cite{KLGB}, one can further reduce any remaining fluctuations
by fitting the values of $\sigma(q)$ (for $q$ big enough) to a straight
line. The
slope of the line should equal $d_{f}$, and the intersection
with the $q$-axis gives $\tau-1$.

We tested this method of analysis for the last wave.
We performed extensive simulations for Sierpinski gaskets
with $5 \leq n \leq 9$. From the statistics of the last
waves we could then estimate $\sigma_{lw}(q)$. The results
are shown in figure 4.
The curvature at low $q$ is a finite size effect. From a fit
of our data in the regime $q \geq 1.5$ we obtain the
estimate $\tau_{lw}=1.28 \pm .03$. This is very close to
the exact result which we obtained in section 2
\footnote{This value of $\tau_{lw}$ is also much lower
then that obtained in \cite{FD2} from a direct fit to
(\ref{8}). We are currently performing a  multifractal analysis of data for the
size and the area of
avalanches in the BTW (and in a stochastic sandpile) model,
on the Sierpinski gasket. The results
will be published elsewhere.}.

\begin{figure}
  \centerline{ \epsfig{figure=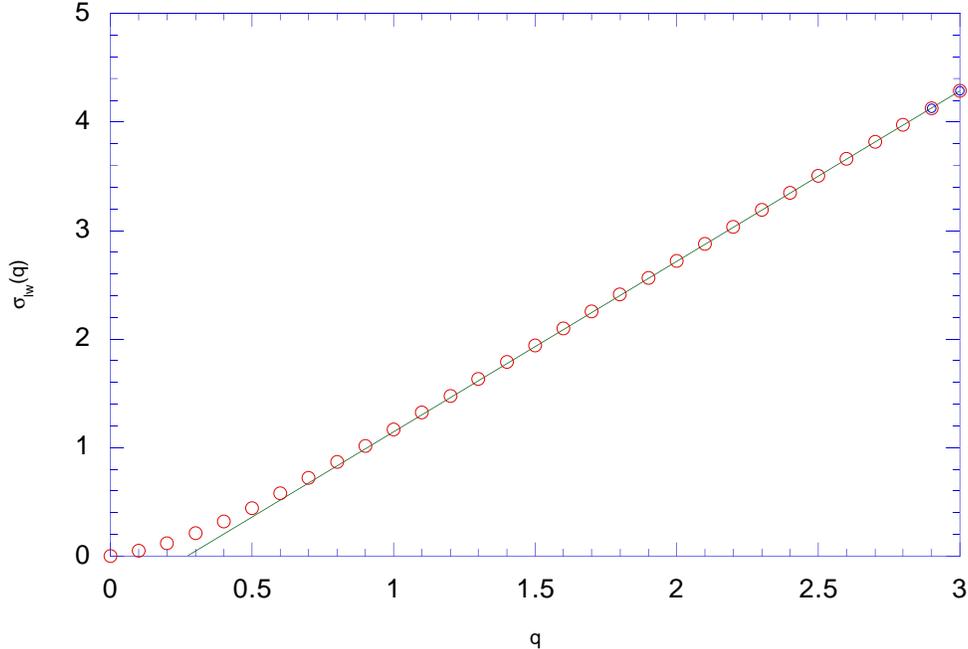,width=13cm}}
\caption{The moment exponent $\sigma_{lw}(q)$ for last waves.
The straight line gives a best fit to the high $q$-data, from
which $\tau_{lw}$ can be obtained.}
\label{figure 4}
\end{figure}

We have followed a similar scheme of analysis for all the other
exponents introduced in section 3. Take as a concrete
example the sum of exponents $\bar{\beta}+\bar{r}$. This can be obtained as
follows. For $s_{k} \gg s_{k+1}$, $P^{+}(s_{k}) \equiv P(s_{k}|s_{k+1}) \sim
s_{k}^{-\bar{r}-\bar{\beta}}$. Instead of analysing this
distribution itself we look at the moments of $P^{+}(s_{k})$ which
are expected to scale as $L^{\sigma^{+}(q)}$. A plot of
$\sigma^{+}(q)$ is shown in figure 5. From an analysis
of the data in this figure we obtain the estimate
$\bar{\beta}+\bar{r}=1.24 \pm .01$, which is
very clearly consistent with the prediction (\ref{3}).

\begin{figure}
  \centerline{ \epsfig{figure=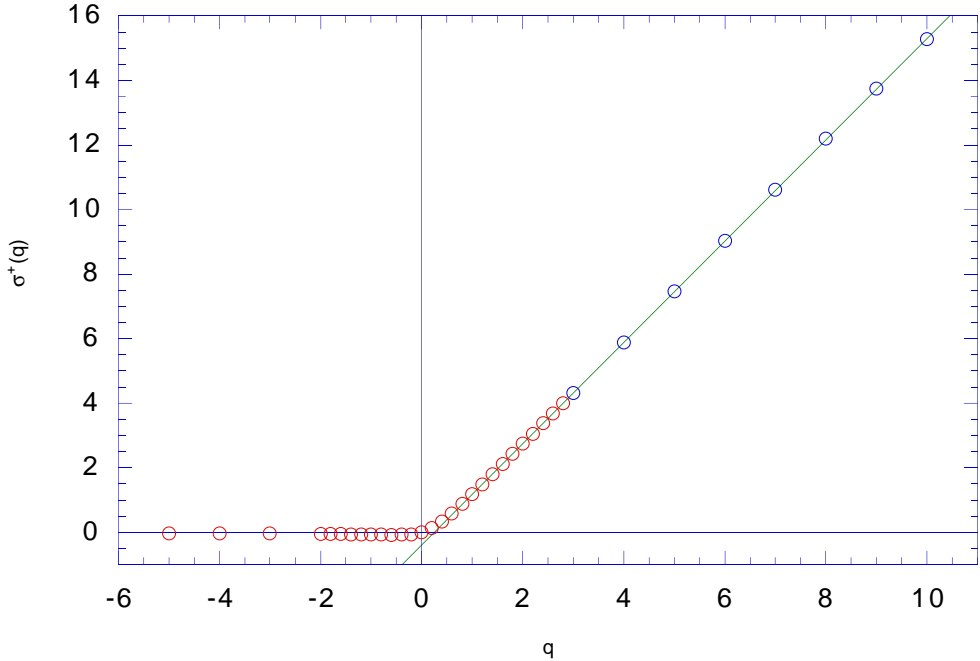,width=13cm}}
\caption{The moment exponent $\sigma^{+}(q)$ (see text).
The straight line gives a best fit to the high $q$-data, from
which $\bar{\beta}+\bar{r}$ can be obtained.} 
\label{figure 5}
\end{figure}

By analysing the small $s_{k}$ behaviour of $P(s_{k}|s_{k+1})$
in a similar way, we can estimate $\bar{\beta}
=.16 \pm 0.05$.

Continuing in this way for the forward conditional probability 
$P(s_{k+1}|s_{k})$, we find from the large $s_{k+1}$-behaviour
$\beta+r=1.30 \pm 0.01$, while from the data for
small $s_{k+1}$ we finally obtain $\beta=0.35 \pm 0.05$
(see figure 6). This numerical result is not too different
from the prediction following from (\ref{14}) which gives $\beta=.288$.

\begin{figure}
  \centerline{ \epsfig{figure=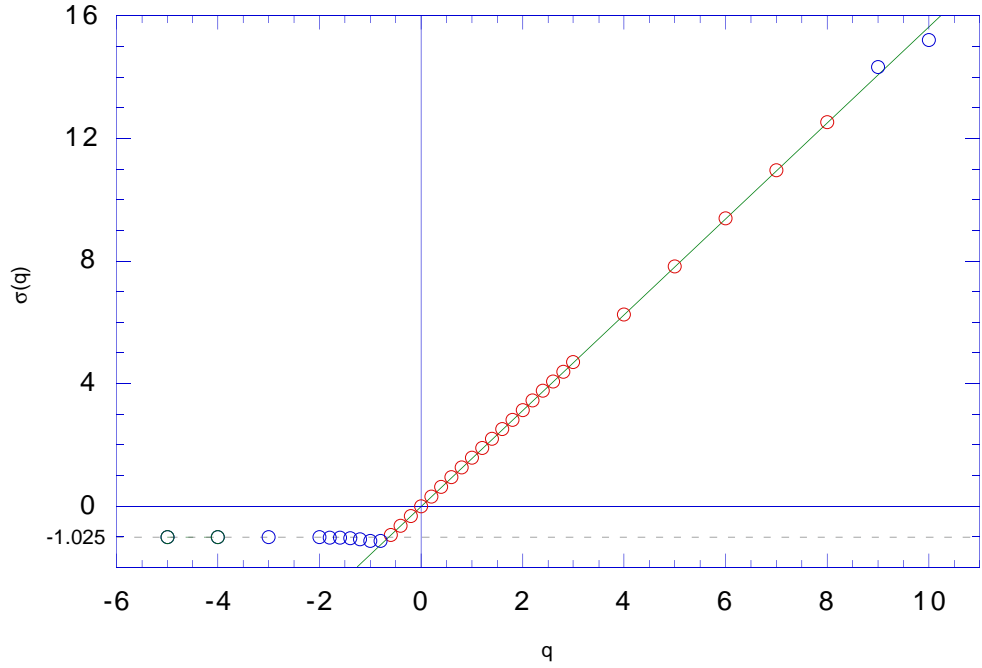,width=13cm}}
\caption{The exponents of the moments of $P(s_{k+1}|s_{k})$ for
$s_{k+1} \ll s_{k}$. The straight line gives a best fit to the
high $q$-data from whicht the exponent $\beta$ can be estimated}

\label{figure 6}
\end{figure}

Finally note that also the relation (\ref{extra.1}) is rather
well satisfied.

\section{Conclusions}
In this paper we investigated the properties of waves in
the sandpile model on a Sierpinski gasket.
We gave predictions for the exponent describing the last
wave in an avalanche and for the scaling exponents occuring
in forward and backward conditional probabilities for
consecutive waves.
These predictions were tested by extensive simulations 
and were found to be in good agreement with the numerics.

Results such as those shown in figure 5 and figure 6
also show no clear evidence for any multifractality
which would show up as a curvature in the plots
of $\sigma(q)$ for big enough $q$ (\cite{Mario2}).

We are currently investigating the presence of
multifractality of avalanches on the Sierpinski
gasket. If, as is the case in $d=1$ (\cite{Kada}) and
in $d=2$ (\cite
{Mario1,Mario2}), such multifractality shows up
, the interesting question arises how such
a phenomenon can be built up on the avalanche
level when it is absent at the level of waves
and consecutive waves.

{\bf Acknowledgement} One of us (VBP) thanks the Limburgs
Universitair Centrum for hospitality.

\end {document}